\documentstyle[prl,aps,amssymb,amsfonts,twocolumn,epsf]{revtex}
\def\1#1{{\bf #1}}
\def\2#1{{\cal #1}}
\def\4#1{{\tt #1}}
\def\5#1{{\sf #1}}
\def\6#1{{\frak #1}}
\def\7#1{{\Bbb #1}}
\def\8#1{{\rm #1}}
\def\9#1{{\cal #1}}
\newtheorem{The}{Theorem}[section]
\newtheorem{Pro}[The]{Proposition}
 \def\Cx{\7C} \def\Ir{\7Z}
\title{\bf Quantum error-correcting codes associated with graphs}
\author{D. Schlingemann and R.F. Werner \\
 {\small Institut f{\"u}r Mathematische Physik, TU Braunschweig,}\\
  {\small Mendelssohnstr.3, 38106 Braunschweig, Germany.}}
\begin{document}
\maketitle
\narrowtext
\begin{abstract} We present a construction scheme for quantum
error correcting codes. The basic ingredients are a graph and a
finite abelian group, from which the code can explicitly be
obtained. We prove necessary and sufficient conditions for the
graph such that the resulting code corrects a certain number of
errors. This allows a simple verification of the 1-error
correcting property of fivefold codes in any dimension. As new
examples we construct a large class of codes saturating the
singleton bound, as well as a  tenfold code detecting 3 errors.
\end{abstract}

\section{Introduction}

From the beginning of Quantum Information Theory it was recognized
that error correcting codes play a crucial role. On the one hand
it was clear that without error correction decoherence effects
could easily annihilate the gain in computing time promised by the
new fast quantum algorithms. On the other hand, the no-cloning
theorem \cite{WoottZur82} seemed to forbid at least the most naive
approach to classical error correction for noisy channels, e.g.,
sending each bit three times and taking a majority vote at the output of the
channel. Clearly, this simple scheme reduces classical errors with
small probability of order $\varepsilon$ to order $\varepsilon^2$.
The cloning required for sending "the same bit" three times rules
out direct quantum analogues of this scheme. It was therefore an
important step to realize \cite{CaSho95}
that quantum mechanics had other, more
subtle ways of ``distributing'' quantum information over several
channels to stabilize against errors. One problem with the known
schemes of quantum error correction
(e.g. \cite{CaRainShoSl96a,CaRainShoSl96b}), however,
is that they tend to
be subtle indeed, and the verification of their error correcting
capabilities often requires a lengthy computation. It is therefore
desirable to find new, perhaps simpler ways of constructing error
correcting codes, on which more direct intuitions might be built.

In this paper we propose a new scheme for constructing quantum
error correcting codes, which has some of these advantages. The
ingredients of our construction are a graph and a finite abelian
group. The order of the group determines the type of systems for
which errors are corrected so that, e.g., the two-element group
corresponds to the qubit case
(Compare \cite{Kn96,Rai97,MatUy99,AshKn00} for other constructions
of non-binary codes).
The number of vertices of the graph
determines how many input systems are coded into how many output
systems. From the edges of the graph one can then read off an
explicit expression for the code. However, not every graph
corresponds to a good code, and we will discuss the condition for
the code to correct a certain number of errors. In the simplest
case, the fivefold code \cite{BeDiViSmoWoo96,LaflMiPaZu96,Chau97b}
(for qubits as well as higher dimensional
systems), it can be verified in a few lines that any two errors are
corrected. We also give an example of a more complex tenfold code
detecting 3 errors.

As we are going to discuss in a following paper in more detail,
it turns out that the codes which can be achieved by our method
are {\em stabilizer codes}. There are
various efficient methods for constructing stabilizer codes
\cite{CaRainShoSl96a,CaRainShoSl96b,Got96,Got97,BeGra99,BeGra99b,BeGeiGra99}.
However, we think that, compared to previous stabilizer constructions,
our technique has some interesting new features.

\begin{itemize}

\item Often the condition for error correction can be proved for
many groups simultaneously, so that one gets {\it code families}
for systems of variable sizes.

\item The geometric intuitions about graphs may become helpful for
finding new constructions.

\item Our codes have the property that in their natural basis all
matrix elements of the coding operator have the same modulus
(Hadamard form). This is prima facie opposite to the usual goal of
getting as few non-zero matrix elements as possible. However, the
latter can be achieved for our codes by a discrete Fourier
transform applied to some channels. Moreover, the Hadamard form
appears to be an interesting normal form for the codes which can
be written in both ways.

\item For some codes it is possible to exchange some input
vertices with some output vertices, while retaining the error
correction property. This kind of symmetry is much harder to see
in the usual stabilizer constructions, and may prove to be helpful
in coding problems with additional inputs and outputs, such as the
internal state of the codig device in convolutional codes.
\end{itemize}

The paper is organized as follows: We begin by describing the
general construction of the coding operator within Section
\ref{sec1}. In Section \ref{sec2} we recapitulate the
Knill-Laflamme condition for error correction and adapt it to our
particular type of codes, resulting in a necessary and sufficient
condition for a graph to generate a quantum error detecting code.
The remaining sections contain examples of codes constructed in this way.
In Section~\ref{ex} we show that it becomes simple indeed to
verify the fivefold quantum codes. In Section~\ref{secsingle}
we demonstrate that for a given number $e$ of errors and number $k$ of
inputs, there is a graph generating an infinite code family using
$4e+k$ output systems, i.e., a family of codes saturating the
singleton bound. Finally, in Section~\ref{tenfolds} we construct a
code with 1 input and 10 outputs, detecting 3 errors for arbitrary system size.

\section{Basic Construction}
\label{sec1}
Every code we construct is completely determined by the follow ingredients:
\begin{itemize}
\item An undirected graph $\Gamma$ with two kinds of vertices.\newline
We distinguish the set $X$ of {\it input vertices} and the set $Y$
of {\it output vertices}. The links of the graph are given by the
{\it coincidence matrix} of the graph, which we will denote by $\Gamma$ for short.
Its matrix element $\Gamma({z_1,z_2})$ is $1$ iff the vertices
$z_1,z_2\in(X\cup Y)$ are linked, and $0$ otherwise. More
generally, we allow {\em weighted graphs}, whose incidence
matrices have arbitrary integer entries, apart from the
constraints $\Gamma({z_1,z_2})=\Gamma({z_2,z_1})$ and
$\Gamma({z,z})=0$.

\item A finite {\it abelian group} $G$ with a non-degenerate symmetric
{\it bicharacter}.
\end{itemize}
By definition, a bicharacter is a function $\chi:G\times G\to\Cx$ such that
$\chi(g+h,g')=\chi(g,g')\chi(h,g')$ and a similar
condition holds for the second argument, which is also implied by the
assumed symmetry $\chi(g,g')=\chi(g',g)$.  We also assume
non-degeneracy in the sense that
\begin{equation}\label{bichar}
  \sum_g \chi(g,g')=|G|\ \delta(g')
  \equiv\left\lbrace\begin{array}{rl}
         |G|&\text{for\ }g'=0 \\
         0  &\text{for\ }g'\neq0 \\
         \end{array}\right.
\end{equation}
Note that since every $g\in G$ has finite order, $\chi(g,g')$ is
always a root of unity, and $\overline{\chi(g,g')}=\chi(-g,g')$.
For $G=\7Z_p$, the cyclic group of order $p$, the standard
bicharacter is given by
\begin{equation}\label{FourierBiC}
  \chi(g,h)=\exp\Bigl(\frac{2\pi i}{p}\; gh\Bigr)\;,
\end{equation}
where $g,h$ are integers representing their class modulo $p$.
Since every finite abelian group is a direct product of cyclic
groups, this also shows the existence of bicharacters for any such
group.

The input and output systems of the code are labeled by $X$ and
$Y$. They are all of the same type, i.e.,  they are described by
the same Hilbert space $\9H=L^2(G)$. This is the space of all
functions $\psi:G\to\Cx$ with scalar product
$\langle\phi,\psi\rangle=|G|^{-1}\sum_g\overline\phi(g)\psi(g)$. For
compactness of notation we write such normalized sums as
integrals. Hence the scalar product becomes $\int\8dg\
\overline\phi(g)\psi(g)$. The combined input system is thus
described in the $|X|$-fold tensor product $\9H^{\otimes
X}=L^2(G^X)$. Vectors in this space are functions of $|X|$
variables, one variable $g_z$ for every $z\in X$. The entire
collection of variables will be denoted by $g^X$. The error
correcting code will be an isometry
\begin{eqnarray}\label{v}
  \1v_\Gamma&:&L^2(G^X)\to L^2(G^Y)\\
  \bigl(\1v_\Gamma\psi\bigr)(g^Y)&=&\int\8dg^X\
  \1v_\Gamma[g^{X\cup Y}]\ \psi(g^X)\;,
\end{eqnarray}
where $\1v_\Gamma$ under the integral denotes the integral kernel of
the operator $\1v_\Gamma$. This kernel depends on both input and output
variables, which are combined into a collection of variables
$g_z$, one for each vertex $z\in X\cup Y$ of the graph.
The core of our construction is an explicit expression for this
integral kernel: apart from an overall normalization factor, it
will simply be a product of phases, with each factor corresponding
to a link of the graph:
\begin{equation}\label{vprod}
  \1v_\Gamma[g^{X\cup Y}]=|G|^{|X|/2} \
     \prod_{\{z,z'\}}\chi(g_z,g_{z'})^{\Gamma(z,z')}\;,
\end{equation}
where the product is over all $2$ element subsets ${\{z,z'\}}\subset(X\cup
Y)$.
Thus for an ordinary graph (${\Gamma(z,z')}=0,1$), this is the product
of all $\chi(g_z,g_{z'})$, for which $z$ and $z'$ are linked. The
remarkable property of such codes is that apart from the
normalization factor the kernel is everywhere of modulus $1$. When
$G=\7Z_p$ is cyclic, and $\chi$ is given by
Equation~(\ref{FourierBiC}), we can write the phase in more
compact form as
\begin{equation}\label{vprodF}
 \1v_\Gamma[g^{X\cup Y}]=|G|^{|X|/2} \
    \exp\Bigl(\frac{\pi i}{p}\;
         g^{X\cup Y}\cdot\Gamma\cdot g^{X\cup Y}\Bigr)\;,
\end{equation}
where ``$\cdot$'' denotes the product of integer valued matrices
and vectors. Note that every term in the sum
$g^{X\cup Y}\cdot\Gamma\cdot g^{X\cup Y}$ occurs twice, which
we compensated by a factor $1/2$.

This completes the construction of the operator $\1v$ from the defining
ingredients listed at the beginning of this section. Of course,
in general this will not be an error correcting code nor even an
isometry. The conditions for this will be studied in the following
section.


\section{The condition for error correction}\label{sec2}
A general characterization of quantum error-correcting codes has first been
worked out by E. Knill and R. Laflamme \cite{KnLafl95}. We briefly
review here the main aspects, and adapt the condition the
particular case of codes constructed as in the previous section.
In this theory a quantum code is an isometry $\1v:\9H\to\9K$ from
the ``input Hilbert space'' $\9H$ to the ``output Hilbert space''
$\9K$. Thus an input density operator is transformed by coding
into $\1v\rho\1v^*$, which is a density operator on $\9K$. The
output of the coding is then passed through a noisy channel. The
noise is described by a certain class of errors, which are
represented by a linear subspace $\9E$ of operators on $\9K$. The
channel is thus represented by a completely positive linear map of
the form
\begin{equation}\label{noise}
  \1T(\rho)=\sum_\alpha F_\alpha\rho F_\alpha^*\;,
\end{equation}
 where $F_\alpha\in\9E$. and are chosen such that the output is
always normalized. The isometry $\1v$ is said to be an error
correcting code for $\9E$, if there is a completely positive
``recovery operator'' $\1R$ such that
\begin{eqnarray}
\1R\bigr(\1T(\1v\rho\1v^*)\bigr)&=&\rho
\end{eqnarray}
 for all density operators on $\9H$. By the theory of Knill and
Laflamme \cite{KnLafl95} this is equivalent to the factorization
condition
\begin{eqnarray}\label{KnLa} \
 \langle\1v\psi_1,F_\alpha^*F_\beta\1v\psi_2\rangle
   &=&\omega(F_\alpha^*F_\beta)\ \langle\psi_1,\psi_2\rangle
\end{eqnarray}
where $\omega(F_\alpha,F_\beta)$ is a factor independent of the
arbitrary vectors $\psi_1,\psi_2$. As in much of the literature on
codes we will consider here a specific type of errors, namely
errors happening only on a small number of outputs of the code.
Thus the tensor product structure $\9H^{\otimes Y}=L^2(G^Y)$ of
the output space becomes important. Let $\6A(E)$ denote the set of
all operators on $L^2(G^Y)$, which are {\it localized} in
$E\subset Y$, i.e., which are the tensor product of an arbitrary
operator on $\9H^{\otimes E}$ with the identity on $\9H^{\otimes
Y\setminus E}$. We say that a code {\it corrects $e$ errors}, if
$F_\alpha,F_\beta$ in (\ref{KnLa}) may be chosen arbitrarily in
the linear span of $\bigcup_{|E|\leq e}\6A(E)$. Note that the
operators $F_\alpha^*F_\beta$ appearing in the scalar product
(\ref{KnLa}) can then be localized on arbitrary sets of $2e$
elements and any operator with such localization may be written as
a linear combination of such $F_\alpha^*F_\beta$. It is therefore
convenient to introduce the following terminology: we say that the
code $\1v$ {\it detects the error configuration $E\subset Y$}, if
\begin{equation}\label{KnLadetect} \
 \langle\1v\psi_1,F\1v\psi_2\rangle=\omega(F)\ \langle\psi_1,\psi_2\rangle\;,
\end{equation}
for all $F\in\6A(E)$. Then a code corrects $e$ errors, iff it
detects all error configurations $E\subset Y$ with $|E|\leq 2e$.

We will now adapt these conditions to operators $\1v_\Gamma$ of the
special form (\ref{v}). Consider a fixed error configuration
$E\subset Y$, and let $I=Y\setminus E$. Then if $F$ is
an operator $F$ on $\9H^{\otimes E}$, with
integral kernel $F[g^E,h^E]$, the integral kernel of
$\1v^*F\1v$ is
\begin{eqnarray}
\1v_\Gamma^*F\1v_\Gamma[g^X,h^X]
&=&\int\8dg^E\8dg^I\8dh^E\
\overline{\1v_\Gamma[g^X,g^E,g^I]}\nonumber
\\ \vspace{0.2cm}
&\times&F[g^E,h^E]\1v_\Gamma[h^X,h^E,g^I]\;.
\end{eqnarray}
This must be a multiple of the identity for every choice of $F$.
Choosing, in particular, a rank one operator $F=|g^E\rangle\langle
h^E|$ we find that error detection for the configuration $E$ is
equivalent to the property that the correlation function
\begin{eqnarray}\label{v*vk0}
\1w_{[\Gamma,E]}[g^{X\cup E},h^{X\cup E}]
&:=&
\int\8dg^I \ \overline{\1v_\Gamma[g^X,g^E,g^I]}\nonumber
\\ \vspace{0.2cm}
&\times&
\1v_\Gamma[h^X,h^E,g^I]
\end{eqnarray}
factorizes in the following manner
\begin{equation}\label{v*vk1}
\1w_{[\Gamma,E]}[g^{X\cup E},h^{X\cup E}]
= C(g^E,h^E)\ \delta(g^X-h^X)\;,
\end{equation}
where $\delta(g^X)$ is defined to be $1$ if  $g_x=0$ for all $x\in X$,
and zero otherwise, and  $C(g^E,h^E)$ is a factor independent of
the input variables $g^X,h^X$.

For two subsets $K,L$ of $X\cup Y$
we denote by $\Gamma^K_L$ the group homomorphism
from $G^L$ to $G^K$ which can be derived from the corresponding
submatrix of the incidence matrix $\Gamma$ by the prescription
\begin{equation}
\Gamma^K_L g^L:=\biggl(\sum_{l\in L}\Gamma(k,l)g_l\biggr)_{k\in K} \ \
\end{equation}
and the following condition is necessary and
sufficient for quantum error detection:

\begin{The}\label{crit1}
Given a finite abelian group $G$ and a  weighted graph $\Gamma$ as in the basic
construction. Then an error configuration $E\subset Y$
is detected by the quantum code $\1v_\Gamma$ iff the system of
equations
\begin{equation}\label{sysequ}
\Gamma^{I}_{X\cup E}d^{X\cup E}=0
\end{equation}
with $I=Y\setminus E$ implies that
\begin{equation}\label{qecccond}
d^X=0 \ \mbox{ and } \ \Gamma^X_E d^E=0  \;.
\end{equation}
\end{The}

The proof of Theorem \ref{crit1} is given in Appendix \ref{app3}
and, in view of this result, we discuss some examples.

\section{Example: The Fivefold Way} \label{ex}

The first example of an optimal quantum error correcting code correcting
{\it all} one bit errors was the famous five qubit code by
\cite{LaflMiPaZu96}. The original code is not easy to verify, so it is
gratifying to see that our construction produces such a code,
which can be verified in a few lines.  Moreover, our construction
works simultaneously for all groups $G$, and is hence not
restricted to qubits. Fivefold codes for higher dimensional
systems have been constructed before \cite{Chau97b}, and if we
believe a recent result by Rains \cite{Rai97c}, the qubit code is
essentially unique anyway. Hence this section has mainly illustrative character.

\begin{figure}[h]
\begin{center}
\epsfxsize=2cm \epsffile{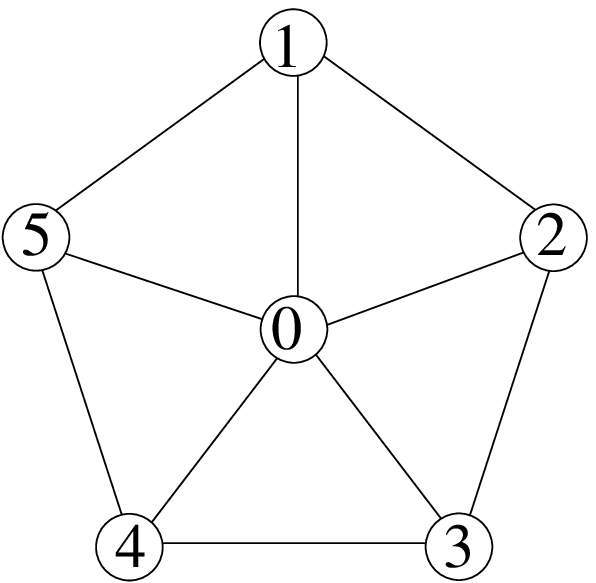}
\epsfxsize=2cm \epsffile{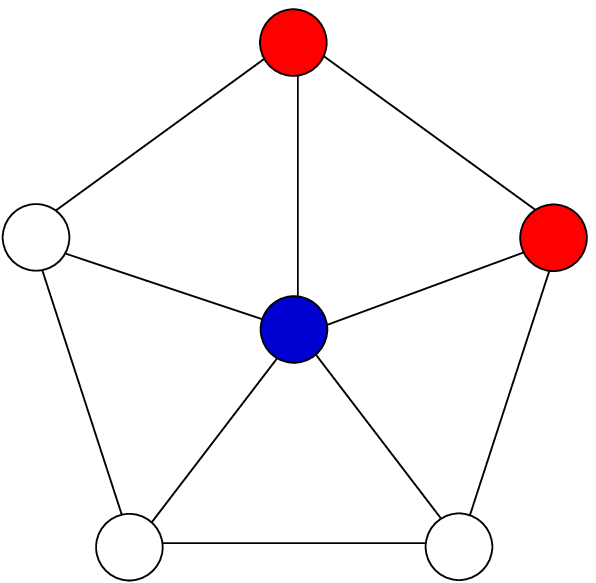}
\epsfxsize=2cm \epsffile{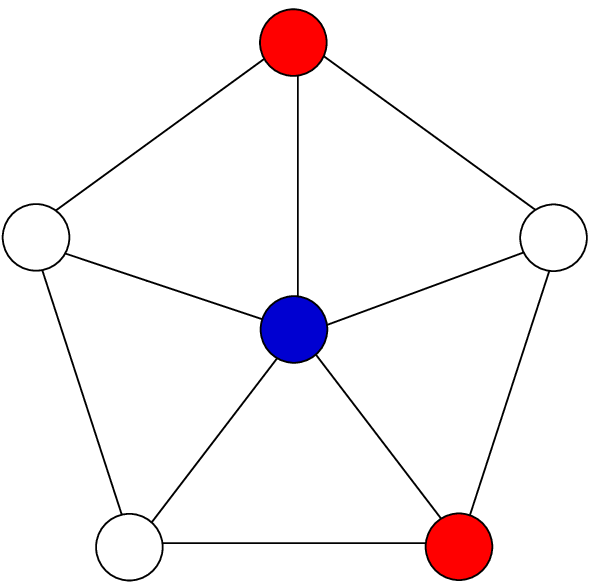}
\end{center}
\caption{The left figure shows a graph for a fivefold code,
where the central vertex is the input vertex. The
two figures on the right side are the relevant 2-error configurations.}
\label{fig1}
\end{figure}

Consider the graph in Figure~\ref{fig1} where the central vertex
"0" is the input vertex, and the remaining five are the output
vertices.
\newpage

We will verify the condition of Theorem~\ref{crit1} in a
particularly strong form. Namely we will show that, for every
2-element error configuration $E$,
 \begin{equation}\label{strongCrit}
   \Gamma^{I}_{X\cup E} d^{X\cup E}=0
\qquad\text{implies}\qquad
 d^{X\cup E}=0\;.
 \end{equation}
The error configuration is a 2-element subset of the output
vertices $1,...,4$, and for the purpose of verifying
criterion~(\ref{strongCrit}) the input vertex 0 plays exactly the
same role as an error.  It is clear by symmetry that only the two
configurations for $X\cup E$ shown by filled dots in
Figure~\ref{fig1} need to be considered. Now the condition
$\Gamma^{I}_{X\cup E} d^{X\cup E}=0$ is a set of equations, one for
each ``integration vertex'' $y\in I$: For each vertex $y$ we have
to sum the $d_x$ for all vertices of $x\in X\cup E$ connected to
$y$, and equate to zero. (In a weighted graph, we would have to
sum with coefficients given by the matrix $\Gamma$). The following
is a table of equations arising in this way for the first error
configuration, $X\cup E=\{0,1,2\}$:
\par\vspace{0.2cm}
\begin{center}
\begin{tabular}[t]{|c|r|}
 \hline
 {\bf Vertex $y$}&{\bf Equation}\\\hline
   3\qquad&$d_0+d_2=0$\\
   4\qquad&$d_0=0$\\
   5\qquad&$d_0+d_1=0$\\\hline
 \end{tabular}
\end{center}
\par\vspace{0.5cm}
\noindent Clearly, this implies $d_0=d_1=d_2=0$ in any abelian
group. Similarly, for the second error configuration $X\cup
E=\{0,1,3\}$we get the equations
\par\vspace{0.2cm}
\begin{center}
\begin{tabular}[t]{|c|r|}\hline
 {\bf Vertex $y$}&{\bf Equation}\\\hline
   2\qquad&$d_0+d_1+d_3=0$\\
   4\qquad&$d_0+d_3=0$\\
   5\qquad&$d_0+d_1=0$\\\hline
\end{tabular}
\end{center}
\par\vspace{0.5cm}\noindent
which once again implies
$d_0=d_1=d_2=0$. This concludes the verification that the code
associated with the graph in Figure~\ref{fig1}, and an arbitrary
finite abelian group $G$, detects any two errors, hence corrects
one error.

In fact, we proved a little bit more than that. The essential part
of the proof was to look at certain $3\times3$-submatrices of the
$6\times6$-matrix $\Gamma$, namely those corresponding to an
off-diagonal block in the partition of the vertices
$\{0,1,\ldots,5\}$ into two disjoint subsets $X\cup E$ and $I$,
and to show that each such submatrix is non-singular. Regarded in
this way, it becomes irrelevant to which of the two sets in the
partition the input vertex ``$0$'' belongs, so we showed that {\it
any} vertex, even a peripheral one, may be taken as input vertex,
and we still get a 1-error correcting code.

\begin{figure}[h]
\begin{center}
\epsfxsize=4cm \epsffile{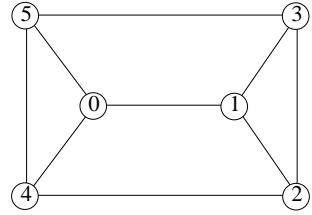}
\end{center}
\caption{An alternative graph for a fivefold code which corrects
one error.}
\label{gr51b}
\end{figure}

This may seem like a rather strong property of the graph we chose.
However, there is (exactly) one other graph with six vertices,
which produces in the same way a 1-error correcting code for
arbitrary choice of input vertex and abelian group $G$. This is
shown in Figure\ref{gr51b}.

\section{Codes saturating the singleton bound}\label{secsingle}

In this Section we briefly discuss one natural generalization of
the idea emerging in the previous section. For definiteness, let
us fix $G$ as a cyclic group of prime order $d$, so $\Ir_d$ is a
field, and choose an integer $m$. We then ask for symmetric
$2m\times 2m$-matrices $\Gamma$ with integer entries (or,
equivalently, entries in $\Ir_d$) with the following property: for
any $m$-element subset $I\subset\{1,\ldots,2m\}$ the $m\times
m$-submatrix $\Gamma_{ij}$ with $i\in I, j\notin I$ is invertible
in the field $\Ir_d$. For the purpose of this section, let us call
such a matrix {\it strongly error correcting} for the prime number
$d$.

What codes can we get from such matrices? Just as in the previous
section, let us specify any set $X$ of $k<m$ vertices as input
vertices, and call the remaining $n=(2m-k)$ ones output vertices.
Then for any configuration of $e=m-k$ errors, the set $X\cup E$
and its complement $I$ will have exactly $m$ elements. By
assumption, the strong form of the error correcting
condition~(\ref{strongCrit}) is satisfied, hence the code detects
$e$ errors. These parameters satisfy
\begin{equation}\label{singleton=}
  n=2e+k\;,
\end{equation}
i.e., the general inequality $n\leq2e+k$, known as the {\it singleton
bound} \cite{KnLafl95} is satisfied with equality for any such code.

How can one get strongly error correcting matrices $\Gamma$ in a
practical way? Here is a procedure, we found  easy to work with for small $m$, using a
symbolic algebra program. First, we introduce variables for each
matrix element $\Gamma_{ij}$ with $i>j$, and compute the
determinants of all off-diagonal $m\times m$-submatrices as
symbolic expressions in these variables. As we go along fixing
integer values for these $\Gamma_{ij}$, the determinant
expressions become simpler, and in some cases factorize. Each of
these factors has to be kept non-zero by the next choice of a
$\Gamma_{ij}$-value. Finally, we end up with an integer matrix,
whose off-diagonal $m\times m$-submatrices all have non-zero
integer determinant. Then, for any prime $d$, which does not
divide any of these integers, we have solved the problem.

\begin{figure}[h]
\begin{center}
\epsfxsize=4.5cm
\epsffile{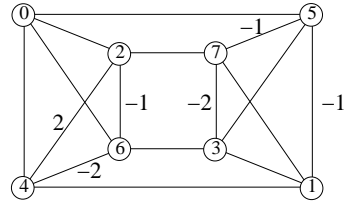}
\end{center}
\caption{Graphical representation of the weighted graph
in Equation (19). Edges without label have weight $1$.}\label{gr71}
\end{figure}

It is natural to begin this process by setting as many weights as
possible equal to zero. It is easy to see that Theorem~\ref{crit1}
does not allow too many $\Gamma(k,l)=0$ because an entire row of
zeros in the matrix $\Gamma^I_{X\cup E}$ leaves one of the
difference variables unconstrained. Similarly, in the condition
for strong error correction it is clear that no off-diagonal
submatrix should have a row of zeros, i.e., each one of the $2m$
vertices must be connected to at least $m$ other vertices. The
graph in Figure~\ref{gr71} is as sparse as possible under these
constraints ($m=4$), and was the starting point for a search for
non-zero weights, as described above, resulting in the matrix

\begin{equation}
\label{m4}
  \Gamma=\left(\begin{array}{cccccccc}
               0&0&1&0&1&1&1&0\\
               0&0&0&1&1&1&0&1\\
               1&0&0&0&2&0&-1&1\\
               0&1&0&0&0&1&2&-2\\
               1&1&2&0&0&0&-2&0\\
               1&1&0&1&0&0&0&-1\\
               1&0&-1&2&-2&0&0&0\\
               0&1&1&-2&0&-1&0&0
          \end{array}\right) \ \ .
\end{equation}
This matrix can either be used to get codes detecting three errors on
an arbitrarily chosen single input vertex, or as a code detecting
two errors (or correcting 1) on two arbitrarily chosen
inputs. The set of determinants is $\{-11, -8, -5, -4, -2, -1, 1,
2, 4, 5, 8, 9\}$, so this will work for any prime $d$ not in in
the set $\{2,3,5,11\}$.
By fixing the choice of the input vertices,
we may restrict to a smaller set of
partitions (the input vertices always belong to the same set),
hence we get fewer constraints. For example, the code with input
vertices $\{1,2\}$ has no relevant subdeterminant containing a
factor $3$, so the resulting code corrects one error on arbitrary
pairs of $3$-level systems.

Within the above example, the number of matrix elements set to zero
is maximal, namely $\nu_i=4$ for each row $i=1,\cdots,8$.
Looking at the corresponding graph, $\nu_i=4$ is just the
number of lines meeting a particular vertex as one can see
from Figure \ref{gr71}.

Strongly error correcting matrices exist in any dimension, so the
code parameters saturating the singleton bound (\ref{singleton=})
can be chosen arbitrarily, if the dimension $d$ of the one-site
system is taken to avoid a certain finite set of primes
(Compare also \cite{Kn96}).
The
argument is quite simple: consider the $m\times m$-
subdeterminants of symmetric $2m\times2m$-matrices as a family of
polynomials $f_\alpha$, $\alpha=1,\cdots,{2n\choose n}$ in ${2n
\choose 2}$ variables. None of these vanishes identically
and since $\7Z$ is an integral domain,
which in contrast is wrong for finite
fields, the product polynomial $\prod_\alpha f_\alpha\not=0$ is non
zero \cite[p. 106]{Jacob}.
This implies that there exists an integer tuple $\Gamma$ of
arguments that $\prod_\alpha f_\alpha(\Gamma)\not=0$.
Thus we have shown the following statement:
\begin{Pro}\label{singleton}
For each number $e$ of errors, there exists a prime $d$
and a weighted graph $\Gamma$ such that the
quantum code, associated with the weighted graph $\Gamma$,
is a quantum error correcting code,
which encodes $k$ $d$-level systems into $4e+k$
$p$-level systems, and which corrects $e$ errors.
\end{Pro}

\section{A 10-fold quantum error-detecting code}\label{tenfolds}

\begin{figure}[h]
\begin{center}
\epsfxsize=4cm \epsffile{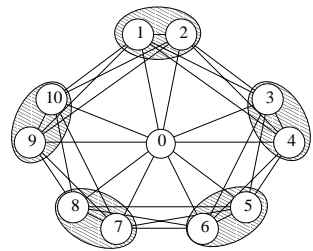}
\end{center}
\caption{The graph for a tenfold code.} \label{ec0}
\end{figure}

In this section we present more complex example for
a graph, which yields for every finite abelian group $G$ a 10-bit
code, detecting 3 errors is given by Figure \ref{ec0}. At the first
look, this graph looks rather complicated, but it can be described
in a simple fashion, by looking at the graph for the fivefold
code in Figure \ref{fig1}.

Namely, the graph, given by Figure \ref{ec0}, can be obtained
as follows:
The output vertices $\{1,2,3,4,5\}$
of the graph in Figure \ref{fig1} are replaced by pairs:
$1\mapsto \{1,2\}$, $2\mapsto\{3,4\}$, \dots ,
$5\mapsto\{9,10\}$.
Each output vertex is connected with the following vertices:
The central input vertex $0$, the vertex belonging the same pair,
and all output vertices belonging to neighbored pairs.

\begin{figure}[h]
\begin{center}
\epsfxsize=3cm\epsffile{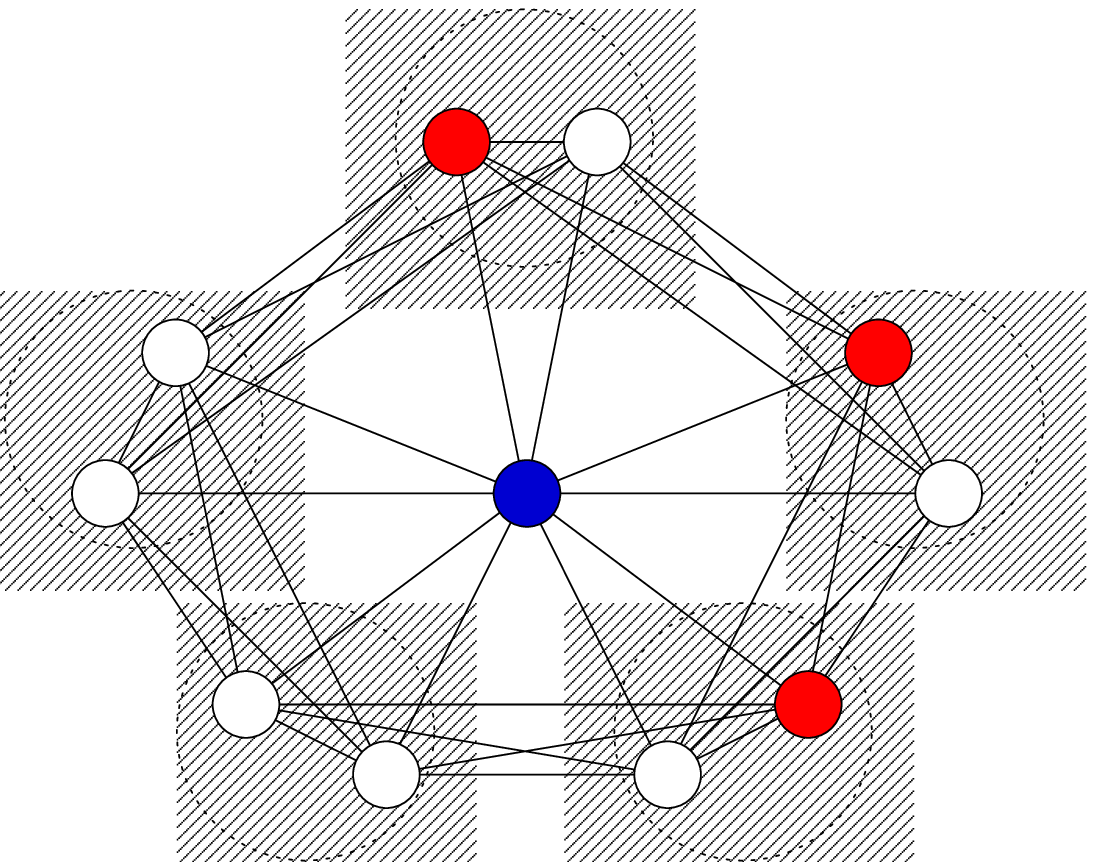}
\epsfxsize=3cm\epsffile{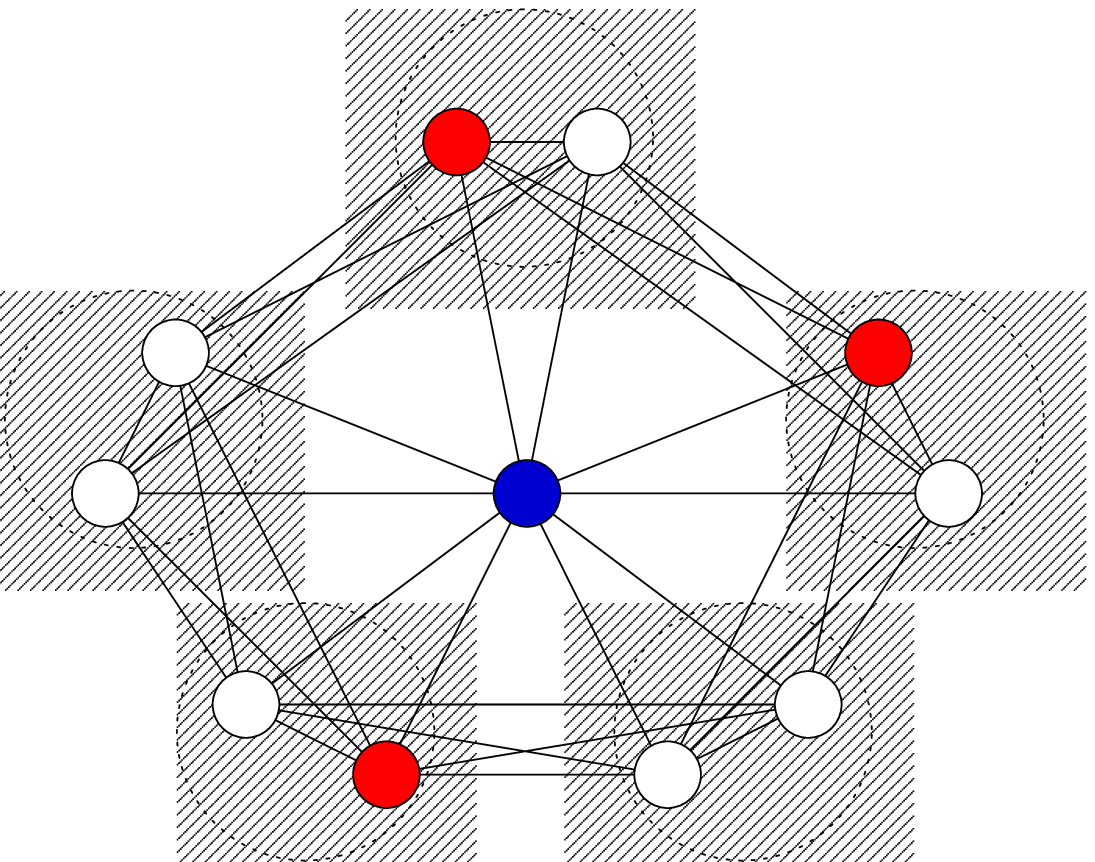}
\epsfxsize=3cm\epsffile{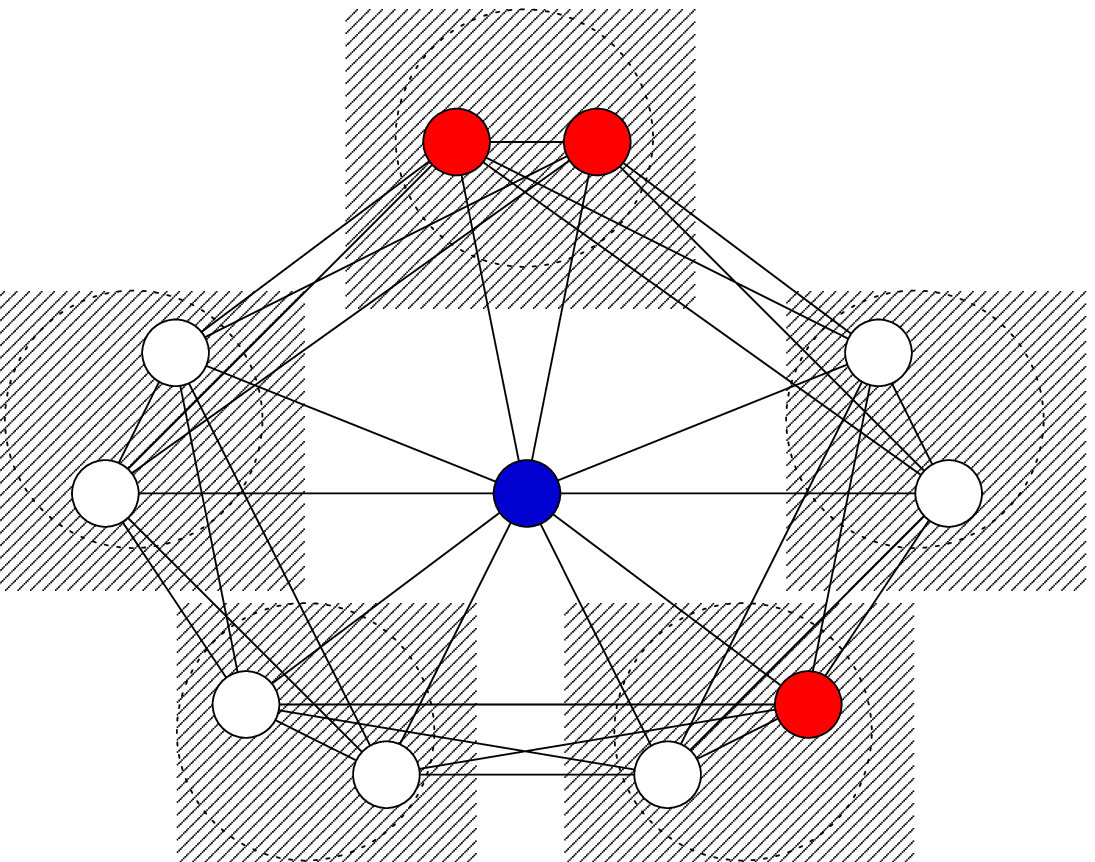}
\epsfxsize=3cm\epsffile{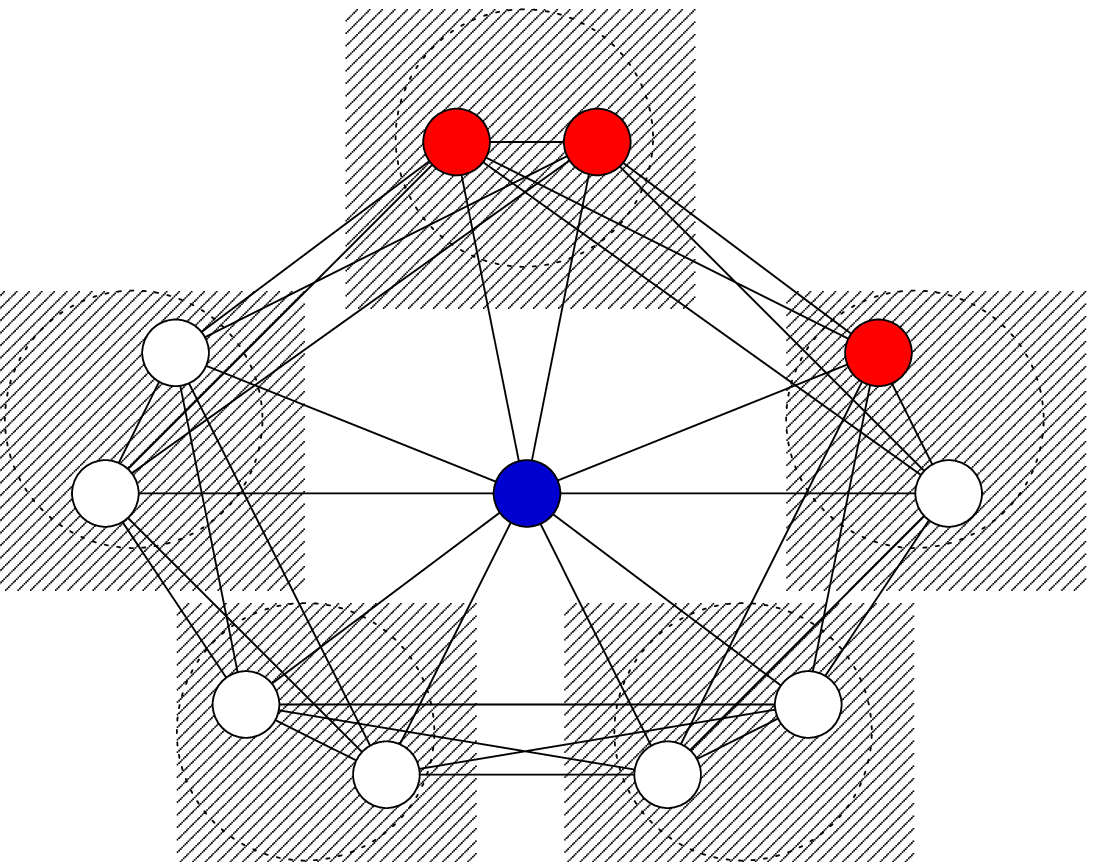}
\end{center}
\caption{A complete family of inequivalent error configurations
for the graph in Fig. 3.}
\label{ec}
\end{figure}

The symmetry of this graph can efficiently be used to check that
each error configuration with 3 errors can be detected. As it is
depicted by Figure \ref{ec}, there are only four 3-error configurations
to distinguish:

\begin{itemize}
\item
All errors occur within different pairs and all these pairs are
neighbored (first graph in Figure \ref{ec}).
\item
All errors occur within different pairs and only two of these pairs are
neighbored (second graph in Figure \ref{ec}).
\item
One pair is totally affected by errors and the remaining error
occurs
within a pair which is not neighbored (third graph in Figure \ref{ec}).
\item
One pair is totally affected by errors and the remaining error
occurs
within a neighbored pair (4th graph in Figure \ref{ec}).
\end{itemize}

\begin{Pro}\label{ind110}
For each finite abelian group $G$,
the quantum code, which is associated with the
graph given by Figure \ref{ec},
is a quantum error-detecting code, encoding 1 input system
into 10 output systems and detecting 3 errors.
\end{Pro}
{\bf Proof:}
Suppose that each error occurs in different pairs and all these pairs are
neighbored (first graph in Figure \ref{ec}), e.g.
the error configuration $\{1,3,5\}$.
Then we proceed as in the example for the fivefold code to
obtain the system of equation (\ref{sysequ}):
\par\vspace{0.2cm}
\begin{center}
\begin{tabular}[t]{|c|r|}
\hline
 {\bf Vertices $y$}&{\bf Equation}\\\hline
   2\qquad&$d_0+d_1+d_3=0$\\
   4\qquad&$d_0+d_1+d_3+d_5=0$\\
   6\qquad&$d_0+d_3+d_5=0$\\
   7 and 8\qquad&$d_0+d_5=0$\\
   9 and 10\qquad&$d_0+d_1=0$\\\hline
\end{tabular}
\end{center}
\par\vspace{0.5cm}\noindent
and we conclude that
$d_0=0$ and $d_1+d_3+d_5=0$ are true. Thus
the equations (\ref{qecccond}) are
fulfilled and the corresponding error configuration is detected.

Analogously, one checks that each error configuration, where each error
occurs in a different pair and only two of these pairs are
neighbored (second graph in Figure \ref{ec}) is also detected.

We now consider an error configuration, where
one pair $\{1,2\}$ is totally corrupted
and the remaining error vertex, e.g. $5$, is contained
within a non-neighbored pair (third graph in Figure \ref{ec}).
We obtain for the system of equations (\ref{sysequ}):
\par\vspace{0.2cm}
\begin{center}
\begin{tabular}[t]{|c|r|}
\hline
 {\bf Vertices} $y$&{\bf Equation}\\\hline
   3 and 4\qquad&$d_0+d_1+d_2+d_5=0$\\
   6, 7 and 8\qquad&$d_0+d_5=0$\\
   9 and 10\qquad&$d_0+d_1+d_2=0$\\
   7 and 8\qquad&$d_0+d_5=0$\\\hline
\end{tabular}
\end{center}
\par\vspace{0.5cm}\noindent
and we conclude again that
$d_0=0$ and $d_1+d_2+d_5=0$ is true.
The equations (\ref{qecccond}) are
fulfilled which implies that the corresponding error
configuration is also detected.

Finally, it is a bit more straightforward as in the previous case
to show that an error
configuration, where two errors occur within one pair and
the remaining error within a neighbored pair (4th graph in Figure \ref{ec})
is detected.
$\square$

\section{Conclusion and outlook}

In this paper we had to limit the exploration of our construction
scheme to a few examples. A more systematic investigation is, of
course, under way. Some of the issues in this investigation are
the following.

\begin{itemize}
\item We already mentioned that all codes constructed from graphs
are stabilizer codes. This is verified by explicitly constructing
a group of unitaries, composed out of shifts and multiplication by
characters, leaving all vectors in the code space invariant. The
converse of this statement is not so clear, i.e., how to embed the
usual stabilizer code constructions into our scheme, and to
characterize the subset of codes for which this is possible.

\item We have seen in Sec.~\ref{ex} that different graphs generate
5-qubit codes, although such codes are presumably unique up to
local transformations. It would be helpful to characterize the
local unitary transformations taking one graph code into another,
and to study the relationships between the resulting graphs. The
Rains invariants \cite{Rai96a,Rai96b,Rai97b}
for graph codes can be computed
relatively easily, and should help to decide such isomorphism
issues.

\item From Sec.~\ref{secsingle} it is clear that the singleton
bound becomes easier and easier to satisfy as the dimension $d$ of
the single system Hilbert space increases. This suggests the
search for bounds describing the resource limitations in coding
more adequately, perhaps by taking into account more detailed
features of the error syndromes than just the maximal number of
errors. Our construction could be helpful for developing and
testing such bounds.

\item Non-stabilizer quantum codes can be constructed from
families of stabilizer codes by taking their {\em union}
\cite{GraBe97c}, where
one has to require that the protected subspaces, corresponding to
the codes within the family, are mutually orthogonal and that this
property remains valid after error operations.
Examples of such non-stabilizer codes are given in
\cite{RaiHarShorSl97,Rusk00}. In view of our construction scheme,
it would be desirable to find sufficient conditions for
a family of graphs such that the union of their corresponding
graph codes yields a (possibly more efficient) non-stabilizer
code.

\end{itemize}
\subsubsection*{{\it Acknowledgment:}}
We would like to thank Mary Beth Ruskai for helpful discussions and
for supporting this investigation with many ideas.
Funding by the European Union project EQUIP (contract IST-1999-11053) is
gratefully acknowledged.
\begin{appendix}

\section{}
\label{app3}
{\bf Proof of Theorem~\ref{crit1}:} We first compute the function
$\1w_{[\Gamma,E]}$, defined by (\ref{v*vk0}). It is convenient to
introduce for two subsets $K,K'$ of $X\cup Y$ the expression
\begin{equation}\label{chiGam}
\chi^\Gamma(g^K,g^{K'}):= \prod_{\{k,k'\}:k\in K,k'\in
K'}\chi(g_k,g_{k'})^{\Gamma(k,k')}
\end{equation}
where the product is taken over all two elementary sets
with one element taken from $K$ and the other taken from $K'$.
Hence the factor for $\{k,k'\}=\{k',k\}$ only occurs once within the
product.
Now we write the
integrand in (\ref{v*vk0}) as a product of two terms
\begin{eqnarray}
\label{decomp}
&&\hskip-25pt
\overline{\1v_\Gamma[g^X,g^E,g^I]}\1v_\Gamma[h^X,h^E,g^I]
\\ \nonumber
&=&|G|^{|X|}{\chi^\Gamma(h^{X\cup E},h^{X\cup E})\over
\chi^\Gamma(g^{X\cup E},g^{X\cup E})} {\chi^\Gamma(h^{X\cup
E},g^{I})\over \chi^\Gamma(g^{X\cup E},g^{I})} \ \ .
\end{eqnarray}
 Here only the last factor on the right hand side depends on the
integration variables $g^I$ associated with the set $I=Y\setminus
E$. In order to carry out the integral over one variable $g_i$,
$i\in I$, we select the $g_i$ dependent part
out of $\chi^\Gamma(h^{X\cup E},g^{I})$, which is
\begin{equation}
\label{part}
\prod_{\{z,i\}:z\in X\cup E} \chi(h_z,g_i)^{\Gamma(z,i)} \ \ .
\end{equation}
Analogously  the $g_i$ dependent part
of $\chi^\Gamma(g^{X\cup E},g^{I})$ is the
same expression with $h$ replaced by $g$.
Thus the $g_i$ dependent part of (\ref{decomp}) is
\begin{equation}\label{part2}
\prod_{\{z,i\}:z\in X\cup E} \chi(g_z,g_i)^{-\Gamma(z,i)}
\chi(h_z,g_i)^{\Gamma(z,i)} \ \ .
\end{equation}
Using the character
property of $\chi(\cdot,g_i)$ we can simplify this to a single
factor of the form $\chi(k,g_i)$. Explicitly,
\begin{equation}
\label{onedelta}
  k=\sum_{j\in X\cup E}\Gamma_{ij}(h_j-g_j)\;,
\end{equation}
 and this sum contains none of the variables associated with $I$, because
$I\cap(X\cup E)=\emptyset$. The integral over $g_i$ then gives
$\delta(k)$, and we find
\begin{eqnarray}
\label{identw}
&&\hskip-15pt
\1w_{[\Gamma,E]}[g^{X\cup E},h^{X\cup E}]
\\
  &=&|G|^{|X|}{\chi^\Gamma(h^{X\cup E},h^{X\cup E})\over
    \chi^\Gamma(g^{X\cup E},g^{X\cup E})}\
    \delta(\Gamma^I_{X\cup E}(h^{X\cup E}-g^{X\cup E})) \ \ .
\nonumber
\end{eqnarray}
Our task is to establish the necessary and
sufficient conditions for this to be of the form
\begin{equation}\label{desired}
 C(g^E,h^E)\ \delta(h^X-g^X)\;,
\end{equation}
 required by Eq.(\ref{v*vk1}).

Now the expression (\ref{identw}) has the required property of
vanishing except for $g^X=h^X$ if and only if this is already
implied by the vanishing of the $\delta$-function in
(\ref{identw}), i.e., if and only $d^X\neq0$ implies
$\Gamma^I_{X\cup E}(d^{X\cup E})\neq0$. This is the first part of
the condition in Theorem~\ref{crit1}.

From now on we assume, as we may,  that
 $\Gamma^I_{X\cup E}(d^{X\cup E})=0$ implies $d^X=0$. Then the
dependence of (\ref{identw}) on the input variables $g^X$ and
$h^X$ can be simplified. The $\delta$-function can be written as
\begin{eqnarray}\label{deltafact}
&&\hskip-30pt
\delta\bigl(\Gamma^I_{X\cup E}(h^{X\cup E} - g^{X\cup E})\bigr)
\\
&=&\delta\bigl(\Gamma^I_{E}(h^{E}-g^{E})\bigr)\ \delta(h^X-g^X)\;,  \nonumber
\end{eqnarray} because the two expressions are equal for $h^X=g^X$, and for
$h^X\neq g^X$ they both vanish by assumption.

To simplify the bicharacter quotient in  (\ref{identw}), we use
 Eq.~(\ref{chiGam}) to write
\begin{eqnarray}\label{threechi}
&&\hskip-30pt
\chi^\Gamma(h^{X\cup E},h^{X\cup E})\\
&=&\chi^\Gamma(h^{X},h^{X})\chi^\Gamma(h^{E},h^{X})
   \chi^\Gamma(h^{E},h^{E})\;. \nonumber
\end{eqnarray}
With a similar decomposition of $\chi^\Gamma(g^{X\cup E},g^{X\cup E})$
we use the condition that, wherever the $\delta$-function in
(\ref{identw}) is non-zero, we have $g^X=h^X$. Hence the $X-X$
factors cancel, and we can write the quotient of the $X-E$-terms
as
\begin{equation}\label{EX-terms}
  {\chi^\Gamma(h^{E},h^{X})\over\chi^\Gamma(g^{E},h^{X})}
   =\prod_{j\in X}\chi\bigl(
        \sum_{i\in  E}\Gamma_{ji}(h_i-g_i),h_j\Bigr).
\end{equation}
For (\ref{identw}) to be of the desired form (\ref{desired}) with
$C(g^E,h^E)$ independent of the $X$-variables, this expression
must be independent of all $h_j, j\in X$, whenever
$\delta\bigl(\Gamma^I_{E}(h^{E}-g^{E})\bigr)\neq 0$. But
(\ref{EX-terms}) is independent of  $h_j$ if and only if
$\sum_i\Gamma_{ji}(h_i-g_i)=0$. Hence we must have that
 $\Gamma^I_{E}(h^{E}-g^{E})=0$ implies $\Gamma^X_E(h^E-g^E)=0$.
This is the second condition from Theorem~\ref{crit1}, which we
have thus shown to be necessary. Conversely, it is sufficient to
ensure  that (\ref{EX-terms}) is equal to $1$, and (\ref{identw})
has the desired form (\ref{desired}) with
\begin{equation}
C(g^E,h^E):=|G|^{|X|}{\chi^\Gamma(h^{E},h^{E})\over\chi^\Gamma(g^{E},g^{E})}
\ \delta(\Gamma^I_E(h^{E}-g^{E})).
\end{equation}
This concludes the proof.
\end{appendix}




\begin{thebibliography}{References}
\bibitem{WoottZur82}
Wootters, W. K. and Zurek, W. H.:
{\it A single quantum cannot be cloned.}
Nature {\bf 299}, 802-803, (1982)

\bibitem{CaSho95}
Calderbank, A.R. and Shor P. W.:
{\it Good quantum error-correcting codes exists.}
Phys. Rev. {\bf A 54}, 1098, (1996)

\bibitem{CaRainShoSl96a}
Calderbank, A.R., Rains, E.M., Shor, P.W., and Sloane, N.J.A.:
{\it Quantum error correction and orthogonal geometry.}
Phys. Rev. Lett. {\bf 78}, (1997), 405-408

\bibitem{CaRainShoSl96b}
Calderbank, A.R., Rains, E.M., Shor, P.W., and Sloane, N.J.A.:
{\it Quantum error correction via codes over GF(4).}
IEEE Transactions on Information Theory, quant-ph/9608006


\bibitem{Kn96}
Knill, E.:
{\it Non-binary unitary error bases and quantum codes}
Los Alamos National Laboratory Report LAUR-96-2717, (1996)

\bibitem{Rai97}
Rains, E.M.:
{\it Nonbinary quantum codes.}
quant-ph/9703048

\bibitem{MatUy99}
Matsumoto, R. and Uyematsu, T.:
{\it Constructing quantum error correcting codes for $p^m$-state
systems from classical error correcting codes.}
IEICE Transactions on Fundamentals of Electronics, Communications
and Computer Sciences, vol. E83-A, no.10, Oct 2000

\bibitem{AshKn00}
Ashikhmin, A. and Knill, E.:
{\it Nonbinary quantum stabilizer codes.}
quant-ph/0005008

\bibitem{BeDiViSmoWoo96}
Bennet, C.H., DiVincenzo, D.P., Smolin, J.A., and Wootters,
W.K.:
{\it Mixed state entanglement and quantum error correction.}
Phys. Rev. {\bf A 54}, 3824, (1996)

\bibitem{LaflMiPaZu96}
Laflamme, R., Miquil, C., Paz, J.-P., Zurek, W.H.:
{\it Perfect quantum error correction code}
Phys. Rev. Lett. {\bf 77}, 198, (1996)

\bibitem{Chau97b}
Chau, H.F.: {\it Five quantum register error correcting
code for higher spin systems.}
quant-ph/9702033

\bibitem{Got96}
Gottesman, D.:
{\it Class of quantum error-correcting codes saturating the quantum
Hamming bound.}
Phys. Rev. {\bf A 54}, 1862, (1996)

\bibitem{Got97}
Gottesman, D.:
{\it Stabilizer codes and quantum error correction.}
PhD thesis (1997)

\bibitem{BeGra99}
Grassl, M. and Beth, T.:
{\it Quantum BCH codes}
Proceedings X. International Symposium on Theoretical Electrical
Engineering, Magdeburg 1999

\bibitem{BeGra99b}
Grassl, M. and Beth, T.:
{\it Cyclic quantum error-correcting codes}
quant-ph/9910061

\bibitem{BeGeiGra99}
Grassl, M., Geiselmann, W., and Beth, T.:
{\it Quantum Reed-Solomon codes}
In Proceedings AAECC-13, 1999

\bibitem{KnLafl95}
Knill, E. and Laflamme, R.:
{\it Theory of quantum error-correcting codes.}
Phys. Rev. {\bf A 55}, 900, (1997)

\bibitem{Rai97c}
Rains, E.M.:
{\it Quantum codes of minimum distance two.}
quant-ph/9704043

\bibitem{Jacob}
Jacobson, N.: {\it Lectures in abstract algebra.}
Vol I. Basic concepts, Springer, New York,  Heidelberg, Berlin, (1964)



\bibitem{Rai96a}
Rains, E.M.:
{\it Quantum weight enumerators.}
quant-ph/9612015

\bibitem{Rai96b}
Rains, E.M.:
{\it Quantum shadow enumerators.}
quant-ph/9611001

\bibitem{Rai97b}
Rains, E.M.:
{\it Polynomial invariants of quantum codes.}
quant-ph/9704042

\bibitem{GraBe97c}
Grassl, M. and Beth, T.:
{\it A note on non-additive quantum codes}
quant-ph/9703016

\bibitem{RaiHarShorSl97}
Rains, E.M., Hardin, R.H., Shor, P.W., and Sloane, N.J.A.:
{\it A non-additive quantum code.}
quant-ph/9703002

\bibitem{Rusk00}
Ruskai, M.B.:
{\it Pauli exchange and quantum error correction.}
quant-ph/0006008




\end{thebibliography}
\end{document}